\documentclass{aa}
\usepackage{natbib}
\usepackage{amssymb}
\usepackage{amsmath}
\usepackage{graphicx}
\def\cdrev#1{#1}
\def\cdrevised#1{#1}
\def\dulrev#1{#1}
\def\dulrevv#1{#1}
\def\cdrfinal#1{#1}
\def\dulfinal#1{#1}
\def\ccline#1{\centerline{#1}}

\def\aap{A\& A}

\def\araa{AnnRevA\& A}

\def\apj{ApJ}
\def\aj{AJ}
\def\apjl{ApJL}

\def\AU{\hbox{AU}}

\begin{document}
\title{Flaring vs.~self-shadowed disks: the SEDs of Herbig Ae/Be stars}
\titlerunning{Flaring vs.~self-shadowed disks}
\authorrunning{Dullemond \& Dominik}
\author{C.P.~Dullemond \& C.~Dominik}
\institute{Max Planck Institut f\"ur Astrophysik, P.O.~Box 1317, D--85741 
Garching, Germany; e--mail: dullemon@mpa-garching.mpg.de\\
Sterrenkundig Instituut `Anton Pannekoek', Kruislaan 403,
  NL-1098 SJ Amsterdam, The Netherlands; e--mail: dominik@science.uva.nl}
\date{DRAFT, \today}

\abstract{Isolated Herbig Ae stars can be divided into two groups (Meeus et
al.~\citeyear{meeuswatersbouw:2001}): those with an almost flat spectral
energy distribution in the mid-infrared (`group I'), and those with a strong
decline towards the far-infrared (`group II'). In this paper we show that
the group I vs.~II distinction can be understood as arising from flaring
vs.~self-shadowed disks. We show that these two types of disks are natural
solutions of the 2-D radiation-hydrostatic structure equations.  Disks with
high optical depth turn out to be flaring and have a strong far-IR emission,
while disks with an optical depth below a certain threshold drop into the
shadow of their own puffed-up inner rim and are weak in the
far-IR. \dulfinal{In spite of not having a directly irradiated surface
layer, self-shadowed disks still display dust features in emission, in
agreement with observations of group II sources. We propose an evolutionary
scenario in which a disk starts out with a flaring shape (group I source),
and then goes through the process of grain growth, causing the optical depth
of the disk to drop and the disk to become self-shadowed (group II source).
We show that this scenario predicts that the (sub-)millimeter slope of the
disk changes from steep (small grains) to Rayleigh-Jeans-like (large grains)
in the early stages of evolution, so that all group II sources are expected
to have Rayleigh-Jeans-like slopes, while some group I sources may still
have steep (sub-)millimeter slopes.}}

\maketitle


\section{Introduction}
Herbig Ae/Be stars are thought to be the intermediate mass counterparts of T
Tauri stars (see e.g.~Waters \& Waelkens \citeyear{waterswaelkens:1998}).
They have a strong infrared (IR) excess arising from warm circumstellar
dust. Similar to T Tauri stars, this dust is believed to reside in a
circumstellar disk. Although there is mounting evidence for the disk-like
distribution of this material, this issue is still not completely
settled. More importantly, the physics and geometry of such disks are not
yet fully \cdrfinal{understood}.

\begin{figure}
\ccline{
\includegraphics[width=9cm]{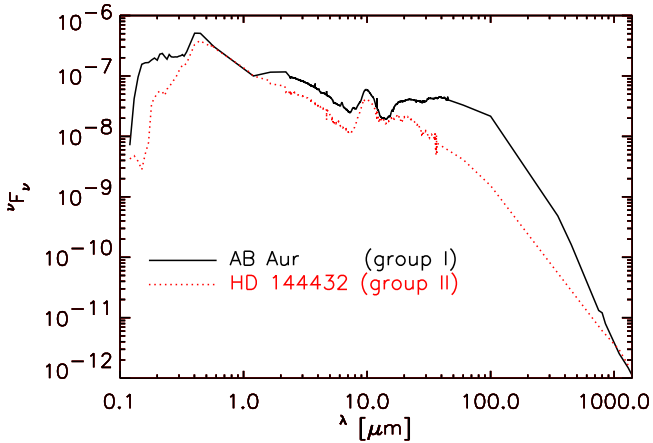}}
\caption{\label{fig-group_i_ii}Two examples of SEDs of Herbig Ae stars of
the sample of Meeus et al. showing the qualitative difference in the SEDs of
group I sources and group II sources.}
\end{figure}

A number of Herbig Ae/Be stars have been studied by Meeus et
al.~(\citeyear{meeuswatersbouw:2001}). They present ISO spectra combined
with ground-based photometry of 13 Herbig Ae/Be stars, and convincingly show
that they can be classified into two main groups, one of which can be
subdivided into two more subgroups. The main division (group I vs.~II)
distinguishes the sources on the basis of the shape of the overall spectral
energy distribution (SED). Group I sources have a relatively strong far-IR
flux, which is energetically comparable with the flux in the near-IR. Group
II sources show a similar near-IR excess as group I sources, but their flux
falls off strongly towards the far-IR.  The fraction of energy reprocessed
by circumstellar dust is typically 30-50\% in group I sources and about 15\%
to 30\% in group II sources.  Examples of these two different kinds of SEDs
are shown in Fig.~\ref{fig-group_i_ii}.

In the paper of Meeus et al.~it was speculated that the distinction between
group I sources and group II sources might be explained qualitatively by the
disk having a flaring geometry or not. \dulrev{A flaring disk captures and
reprocesses more stellar radiation at large radii than flat disks do, and
therefore naturally has a stronger far-infrared excess. However, it remained
unclear what physical mechanism could cause some disks to be flaring and
others to be flat.}   In fact, models of
passive circumstellar disks (e.g.~D'Alessio et
al.~\citeyear{dalessiocanto:1998}; Chiang \& Goldreich
\citeyear{chianggold:1997} CG97; Dullemond, Dominik \& Natta
\citeyear{duldomnat:2001} DDN01) have so far shown that these disks
consistently have a flaring geometry.

\cdrevised{A possible mechanism that could flatten a disk is dust settling.
  Due to the gravity excerted by the central star, dust grains can slowly
  drift towards the midplane of the disk while the gas remains behind.
  Since the stellar radiation is absorbed almost exclusively by dust, the
  photosphere of the disk flattens as dust settling proceeds. This mechanism
  has been proposed by Chiang et al.~(\citeyear{chiangjoung:2001}) as an
  explanation for the mid- and far-IR SEDs of the Herbig stars MWC 480 and
  HD36121. They found that the SEDs of these sources could only be fitted
  with a CG97 type disk if the vertical geometric thickness was artificially
  reduced by a factor of about 3 to 5.} \dulrevv{They argue that this
  suggests that dust settling is the mechanism behind the group II type
  SEDs.}

\dulrevv{In this paper we will demonstrate a more direct, though possibly
  related, way of causing a disk to have a group II type SED.
  It is the effect of {\em self-shadowing}. If the disk's geometric thickness
  $H/R$ is a monotonic increasing function of radius $R$, then the disk is
  flaring, and can capture stellar radiation at all radii. If, on the other
  hand, the ratio $H/R$ becomes smaller as one goes to larger radii, then
  the disk becomes self-shadowed: the inner disk regions cast a shadow over
  the outer disk regions. As the shadow deprives the outer regions from most
  of their irradiation flux (only indirect irradiation remains possible, but
  is rather weak), the far-IR emission is strongly suppressed.
  This works even without the need of a factor of 3 to 5 reduction
  in $H$, as long as the outer parts of the disk are in the shadow of the
  inner parts. Dust settling could be a possible mechanism causing such a
  shadowing (Dullemond \& Dominik in prep.), but another possibility is the
  vertical optical depth of the disk, as we will show in this paper.}

\dulrevv{We will show that an important role in the self-shadowing is
  played by the disk's puffed up inner rim. The existence of such a
  \cdrfinal{structure} was inferred from observations of Herbig Ae/Be
  stars in the near-IR.  Virtually all Herbig Ae/Be stars show a
  prominent thermal bump in their SEDs around 3\,$\mu$m.}  \cdrevised{
  Since a simple flaring disk model is not capable of reproducing it,
  several authors invoke additional non-disk components to explain this
  feature (Miroshnichenko et
  al.~\citeyear{miroiveeli:1997},\citeyear{miroivevinkeli:1999}; van den
  Ancker et al.~\citeyear{vdanckerbouw:2000}).  But Natta et
  al.~(\citeyear{nattaprusti:2001}) have drawn attention to the importance
  of the inner hole in the disk caused by dust evaporation, and the emission
  from the inner rim of the dusty part of the disk.}  \dulrevv{It was shown
  by Dullemond, Dominik \& Natta (\citeyear{duldomnat:2001}, henceforth
  DDN01) that the model of Chiang \& Goldreich, extended by a
  self-consistent (though simplified) description of the inner rim, could
  indeed explain the entire SEDs of Herbig Ae stars, in particular those of
  group I. An important feature of these models is that the inner rim of the
  dusty disk is {\em puffed-up} as a result of its direct exposure to the
  light of the central star.  The rim therefore casts a shadow over part of
  the disk, typically from the location of the rim ($\sim$ 0.5 AU) out to
  about $\sim$ 5 AU. Only further out the surface height of the disk becomes
  sufficiently high to cause the upper layers of the disk to reach out of
  the shadow wedge cast by the inner rim (see Fig.~\ref{fig-pictograms} 
  left).}

\dulrevv{While the DDN01 models work best for group I sources, they can
  also fit group II sources (even without dust settling). But these fits
  require an unlikely termination of the disk just outside of the shadow
  cast by the inner rim (Dominik et al.~\citeyear{domdulwatwal:2003}). }
\cdrevised{It seems unlikely that all group II sources have such a
  fine-tuned outer radius.  For some sources it is also known from other
  observations that the disks are much larger (Mannings \& Sargent
  \citeyear{mannsarg:1997}).}
\dulrevv{Dominik et al.~argue that the inferred smallness of the flaring
  part of the disk may in fact be a hint that these disks are fully
  self-shadowed, i.e.~that the inner rim shadows the entire disk. However,
  no detailed models of such self-shadowed disks were given.}

\cdrevised{A consistent discription of \dulrevv{self-shadowed (non-flaring)}
  disks can only be achieved by employing a 2-D (axisymmetric) or 3-D
  radiative transfer code to compute the temperature structure of the
  disk. By also computing the vertical {\em density} structure of the disk
  simultaneously, it can be investigated if, and under what circumstances
  non-flaring or self-shadowed disks can exist.  A first attempt in this
  direction was done by Dullemond (\citeyear{dullemond:2002}, henceforth
  D02).  Using a grey 2D radiative transfer code it was shown that disks
  with a high optical depth flare in the way described by DDN01 (inner rim,
  shadow and outer flaring part).  But when the optical depth of the disk
  is below a certain threshold, the entire disk drops into the shadow of
  its own inner rim (see Fig.~\ref{fig-pictograms} right).}  \cdrfinal{The
  resulting SED} turns out to be very similar to group II sources: a strong
  near IR bump and a weak far-IR flux with a steep slope.  \dulrevv{In the
  D02 paper it was therefore suggested that the group II type of SED is
  caused by {\em self-shadowing}, instead of a mere flattening of a flaring
  disk.}

\dulrevv{In spite of these new findings,} a number of important questions
remain to be answered before firm conclusions can be drawn.  In particular,
the assumption of grey opacities in D02 is very unrealistic.  If the disk is
dominated by small grains (as indicated by the prominent emission features)
the opacity is far from grey.  It is possible that real opacities will
change the picture considerably.  Moreover, with real opacities we can
compare the model to the observations in more detail, and we can assign
realistic masses to our disk models where previously we could only speak of
the optical depth. Therefore, the present paper addresses the following
issues:
\begin{itemize}
\item Does the group I/II explanation in terms of flaring/self-shadowed
still hold when real opacities are used? If so, for which disk parameters 
do we get flaring or self-shadowed disks? Does dust grain growth play
a role in the division between group I and group II sources?
\item Will self-shadowed disks provide the 10 micron feature in emission, as
seen in most group II sources?  \cdrevised{In the simple picture of a CG97
or DDN01 disk, these features are always formed in the superheated layer
created by dust directly illuminated by the star. In a self-shadowed disk
the disk surface is not directly irradiated.}
\item What do the models predict for the (sub-)mm flux and slope for group
I and group II disks? The grey models were entirely optically thick in the 
(sub)mm, which may not be the case when real opacities are used.
\end{itemize}

In this paper we present the \emph{first self-consistent 2-D models of
Herbig Ae star disks with realistic dust opacities}, and we show that indeed
the discrepancy between group I and group II sources can be understood
naturally in terms of flaring and self-shadowed (i.e.~non-flaring) disks.
We also show that through the reduction of the optical depth of the disk by
the growth of grains, a flaring disk can be turned into a self-shadowed
disk. This suggests a natural evolutionary link between group I and group
II sources.

The paper is organized as follows: in section~\ref{sec:models} we describe
the modeling procedure and radiative transfer codes used.  In
section~\ref{sec:results} we describe the structure of - and the SEDs
produced by - two sets of models which run from large to low optical depth
of the outer disk parts.  In section~\ref{sec:discussion} we discuss the
results in the framework of shadowing, rim emission, solid state features
and grain growth.

\section{Models}
\label{sec:models}

\subsection{Modeling procedure}\label{sec-model-proced}

The equations for the models of this paper are very similar to the ones
presented in D02, but while in D02 a grey opacity is used for the sake of
simplicity, we now use silicate grains with a size of $a=0.1\mu$m (Draine \&
Lee \citeyear{drainelee:1984}). In order not to complicate the models
unnecessarily we ignore the scattering opacity (we set it to zero). An
extensive discussion of the effects of scattering in models of passive
reprocessing disks is given in Dullemond \& Natta
(\citeyear{dulnatscat:2003}).

To compute the disk structure, we proceed as follows. We adopt as our
coordinate system the polar coordinates $R$ and $\Theta$ (where $\Theta=0$
means the pole and $\Theta=\pi/2$ is the equator), and we assume axial
symmetry. The disk is presumed to be passive, i.e.~non-accreting, so that
the energy balance is entirely determined by the irradiation by the central
star. This process of irradiation is modeled using a 2-D axisymmetric
continuum radiative transfer code that solves for the dust temperature as a
function of $R$ and $\Theta$. 

The determination of the density structure requires an iteration procedure.
An input to the model is the surface density distribution $\Sigma(R)$. At
the start we set up a reasonable initial guess for the density distribution
$\rho(R,\Theta)$ that is consistent with this $\Sigma(R)$. We can then apply
the radiative code to find the gas temperature everywhere. Using this
temperature distribution, a vertical integration of the equation of vertical
hydrostatic equilibrium yields the new 2-D density structure
$\rho(R,\Theta)$. After iterating the entire procedure of radiative transfer
and vertical structure a couple of times one finds a solution in which both
$\rho(R,\Theta)$ and $T(R,\Theta)$ are consistent with each
other. \dulrev{Using this density and temperature structure a ray-tracer can
now produce the required SEDs and images.}

\dulrev{There is, however, a major hurdle that needs to be overcome before
the above schetched procedure can work. The disks that we wish to model may
have tremendous optical depths. An optical depth of $\tau_{V}=10^5$ is not
uncommon. Most radiative transfer codes cannot handle such optical depths:
they will either produce noisy or wrong results, or they never converge.  As
was described in D02, to solve this problem we use the code RADICAL, which
is based on the method of {\em Variable Eddington Tensors (VET)}. This
method is the only method presently known to be able to treat problems with
such enormous optical depth without requiring excessive excecution time. For
1+1D type disk models they have been succesfully applied many times already
(see e.g.~Malbet et al.~\citeyear{malbetbertout:1991},
\citeyear{malbetlachaume:2001}; Dullemond, van Zadelhoff \& Natta
\citeyear{dulvzadnat:2002}). For 2-D radiative transfer problems they are
slowly coming in use (e.g.~Nakazato, Nakamoto \& Umemura
\citeyear{naknakume:2003}; D02). }

\dulrev{The problem with the VET algorithm implemented in RADICAL is that it
sometimes has slight inaccuracies in the resulting temperature profiles.
These inaccuracies are usually not larger than about 5\% to 10\%, but since
the SED is proportional to $T^4$, such inaccuracies are amplified in the
outcoming spectrum. For the structure iteration these inaccuracies are less
dangerous, as the pressure scale height of the disk is proportional to
$\sqrt{T}$, meaning that the error in the density structure of the disk is
2.5\% to 5\%. It seems therefore that RADICAL
is appropriate for the structure iteration procedure, but not for computing
the temperature profile for the ray-tracer producing the SED and images.}

\dulrev{For the final temperature structure determination (after convergence
of the density structure) we therefore} prefer to use a different method: an
improved version of the algorithm of Bjorkman \& Wood
(\citeyear{bjorkmanwood:2001}), implemented in a code called {\tt
RADMC}. \dulrev{As expected, this method suffers from large CPU costs at
high optical depth, and in many cases produces an strong numerical noise on
the temperature profile in the most optically thick regions. But this is not
a major problem since these optically thick regions (usually near the
equator at small radii) are unobservable. And since RADMC
is called only once per model (while RADICAL has to be called many times
during the structure iteration), the large CPU cost of RADMC remains
manageable.  } RADMC has the advantage of being very accurate and reliable
when it comes to the emerging spectrum. We have found that for some of the
models RADICAL slightly overpredicts the 5--8 $\mu$m part of the SED, while
RADMC produces a sharper transition between the mid-infrared-bump and the
10-micron feature, in agreement with the SEDs predicted by semi-analytic
models (Dullemond, Dominik \& Natta \citeyear{duldomnat:2001}), and found in
nature (e.g.~Meeus et al.~\citeyear{meeuswatersbouw:2001}).  We verified
that the density structure produced after the RADMC run is virtually
identical to the one produced with the RADICAL code. Also, both codes have been
independently tested against other codes on a number of test cases, albeit
of much less extreme optical depths than here (Pascucci et
al.~\citeyear{paswolstedulhen:2004}). \cdrfinal{Both the RADICAL results and
the RADMC results conserve energy to within an error of at most 5\%.} 

\dulfinal{As final note it should be mentioned that the iteration procedure,
in which the hydrostatic equilibrium code and the radiative transfer code
are alternately applied, may in some cases not reach a perfectly converged
solution. Minor waves may propagate over the solution from one iteration to
the next and never damp out completely. This problem is related to a known
instability operating in these disks (Dullemond \citeyear{dullemond:2000})
and may point to such waves propagating the disk in reality as well. For
disks around Herbig Ae/Be stars, like the models presented in this paper,
these waves are weak and have virtually no effect on the SED or images. For
T Tauri stars the problem is more serious, and a study of the stability of
these disks using time-dependent heating and cooling calculations is
required. For this reason we limit ourselves in this paper to Herbig Ae/Be
stars only.}

\subsection{Setup of the models}
The models of this paper
simulate a disk around a star of $M_{*}= 2.5 M_{\odot}$, $R_{*}=2R_{\odot}$
and $T_{*}=10000 K$. The surface density of the disk as a function of radius
$\Sigma(R)$ is defined to be:
\begin{equation}
\Sigma(R) = \Sigma_0 (R/R_0)^p 
\end{equation}
with $R_0$ taken to be $R_0=200\AU$. The mass of the disk is then:
\begin{equation}
M_{\mathrm{disk}} = \left\{ \begin{matrix}
2\pi  \Sigma_0 R_0^{-p} \frac{1}{p+2} 
 \Big[R_{\mathrm{out}}^{p+2}-R_{\mathrm{in}}^{p+2}\Big] 
& \hbox{(for }p\neq 2\hbox{)} \\[4mm]
2\pi \Sigma_0 R_0^{-p}  \hspace{\fill}
 \Big[ \ln R_{\mathrm{out}}-\ln R_{\mathrm{in}}\Big] 
& \hbox{(for }p= 2\hbox{)} 
\end{matrix}\right.
\end{equation}
\dulrevv{Similar equations hold for the dust mass of the disk
$M_{\mathrm{dust}}$ in relation to the dust surface density
$\Sigma_{\mathrm{dust}}$. In general one has
$M_{\mathrm{dust}}=0.01\,M_{\mathrm{disk}}$ and
$\Sigma_{\mathrm{dust}}=0.01\,\Sigma_{\mathrm{disk}}$, because of the
gas-to-dust ratio of 100.}

We take the inner radius to be at $0.5\AU$, which is approximately the dust
evaporation radius, and we put the outer radius at $200\AU$.  We allow the
disk to extend a bit further than $200\AU$, but with a very steep powerlaw
index for the surface density $\Sigma\propto R^{-12}$, so that effectively
the disk has a is 200 AU outer radius, but does not end too abruptly there.

\dulrevv{In this paper we present three series of models:
\begin{itemize}
\item {\bf Series A:} Disks with equal mass ($M_{\mathrm{disk}}=0.01
M_{\odot}$, i.e.~$M_{\mathrm{dust}}=0.0001 M_{\odot}$), but different
distribution of $\Sigma(R)$ (power law index between $p=-1\cdots -4$). The
grain size is $0.1\mu$m.
\item{\bf Series B:} Disks with varying mass (between
$M_{\mathrm{disk}}=0.1\cdots 10^{-6} M_{\odot}$,
i.e.~$M_{\mathrm{dust}}=10^{-3}\cdots 10^{-8} M_{\odot}$), but with the same
distribution of this mass as a function of radius ($p=-1.5$). The grain size
is again $0.1\mu$m.
\item{\bf Series BL:} Like the B series, but the reduction in mass of the
disk is now a reduction only in the small grain dust mass ($0.1\mu$m).  As
the mass in small grains is reduced, the removed dust mass is converted into a
layer of large (2 mm) grains at the equatorial plane, so that the total dust
mass ($0.1\mu$m and 1 mm grains) remains the same
($M_{\mathrm{dust}}=0.001\,M_{\odot}$). In this way we simulate the process of
grain growth, and the subsequent settling of the large grains to the
midplane. The small grains are assumed {\em not} to have settled.
\end{itemize}
}
The parameters of all these models are
listed in table \ref{table-model-parameters}.

\cdrfinal{The A series allows us to see whether self-shadowed disks can
  result from redistributing matter in the disk from the outside to the
  inside.  The B series shows whether an overall reduction of the optical
  depth (i.e.~mass) can cause self-shadowing. The BL series is like the B
  series, but simulates grain growth and settling (see D'Alessio, Calvet \&
  Hartmann~\citeyear{dalessiocalvet:2001} for an earlier study of the effect
  of grain growth on disk models).}

\begin{table}
\ccline{
\begin{tabular}{c|ccc|c}
   & $p$ & $M_{\mathrm{dust,small}}/M_{\odot}$ & 
  $M_{\mathrm{dust,big}}/M_{\odot}$ & $M_{\mathrm{disk}}/M_{\odot}$ \\
\hline
A1  & -1   & $10^{-4}$ & 0 & $10^{-2}$ \\
A2  & -2   & $10^{-4}$ & 0 & $10^{-2}$ \\
A3  & -3   & $10^{-4}$ & 0 & $10^{-2}$ \\
A4  & -4   & $10^{-4}$ & 0 & $10^{-2}$ \\
\hline                                 
B1  & -1.5 & $10^{-3}$ & 0 & $10^{-1}$ \\
B2  & -1.5 & $10^{-4}$ & 0 & $10^{-2}$ \\
B3  & -1.5 & $10^{-5}$ & 0 & $10^{-3}$ \\
B4  & -1.5 & $10^{-6}$ & 0 & $10^{-4}$ \\
B5  & -1.5 & $10^{-7}$ & 0 & $10^{-5}$ \\
B6  & -1.5 & $10^{-8}$ & 0 & $10^{-6}$ \\
\hline           
BL1 & -1.5 & $10^{-3}$ &        0            & $10^{-1}$ \\
BL2 & -1.5 & $10^{-4}$ & $9.0\times 10^{-4}$ & $10^{-1}$ \\
BL3 & -1.5 & $10^{-5}$ & $9.9\times 10^{-4}$ & $10^{-1}$ \\
BL4 & -1.5 & $10^{-6}$ & $9.99\times 10^{-4}$ & $10^{-1}$ \\
BL5 & -1.5 & $10^{-7}$ & $1.0\times 10^{-3}$ & $10^{-1}$ \\
BL6 & -1.5 & $10^{-8}$ & $1.0\times 10^{-3}$ & $10^{-1}$ \\
\end{tabular}}
\caption{\label{table-model-parameters} \dulrevv{The parameters of the three
series of models presented in this paper.  $p$ is the power law index
for the surface density ($\Sigma(R) \propto R^p$). 
$M_{\mathrm{dust,small}}$ is the mass in small (0.1 $\mu$m) grains. These
grains are evenly distributed with the gas in a constant gas-to-dust mass
ratio. For the A and B models, this ratio is 100. For the BL models this
ratio is 100,1000,10\,000 etc., but still constant over the disk. The
$M_{\mathrm{dust,big}}$ is the mass in big (2 mm) grains which are assumed
to be located in a thin midplane layer. This parameter is only non-zero for
the BL series, and it is taken such that
$M_{\mathrm{dust,small}}+M_{\mathrm{dust,big}}=10^{-3}\,M_{\odot}$. In this
way, the BL series simulates a process of converting small grains evenly
distributed in the disk into big grains located at the midplane.
\cdrfinal{$M_{\mathrm{disk}}$ is the total mass of the disk
(dust+gas) as calculated from the three disk parameters.  It is
therefore not a model parameter.}  Note that the B1
and BL1 models are identical.}}
\end{table}

\section{Results}
\label{sec:results}
\subsection{Structure of the disk}
\label{sec-struct-results}
\begin{figure*}
\ccline{
\includegraphics[width=9cm]{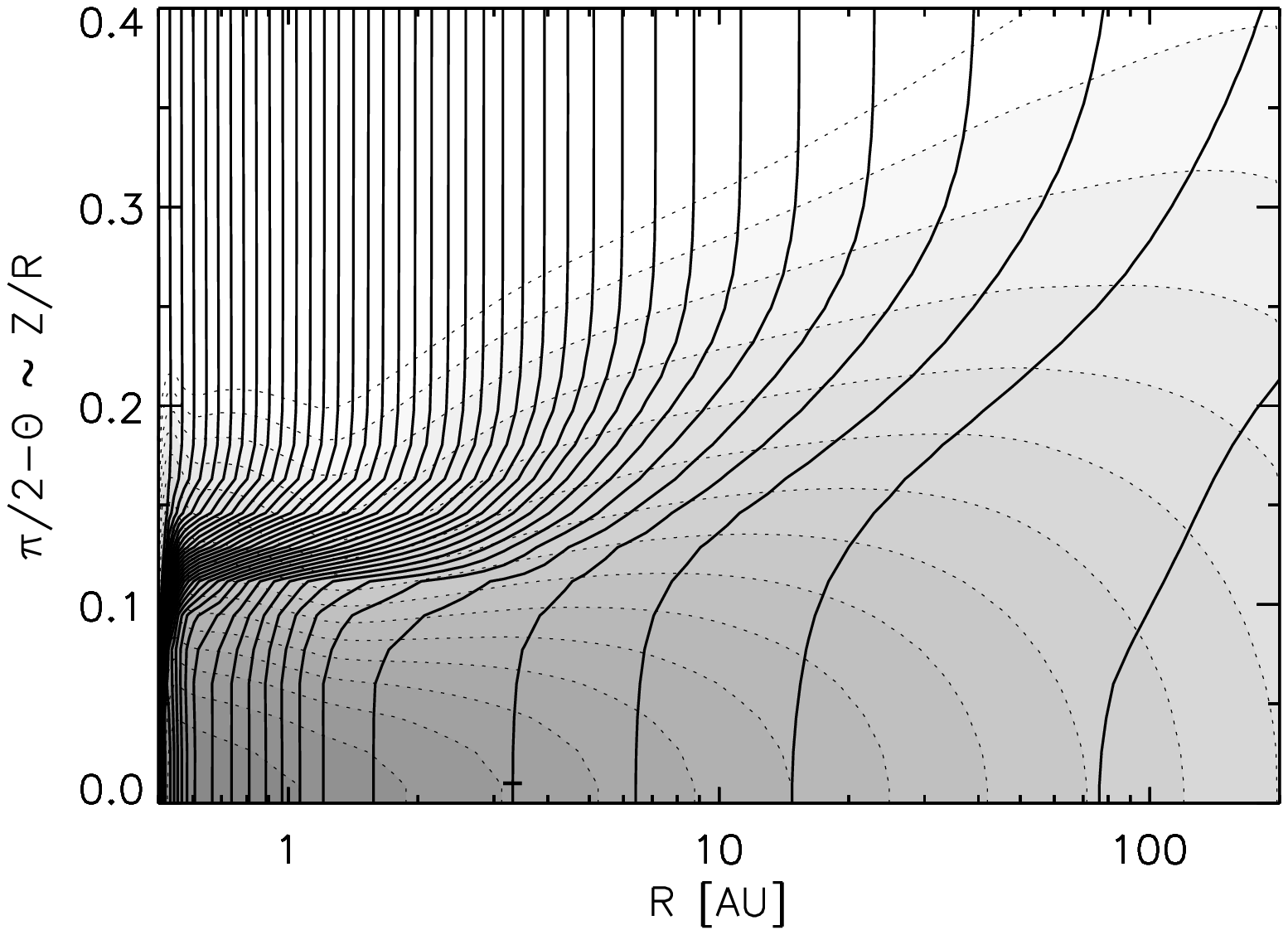}
\includegraphics[width=9cm]{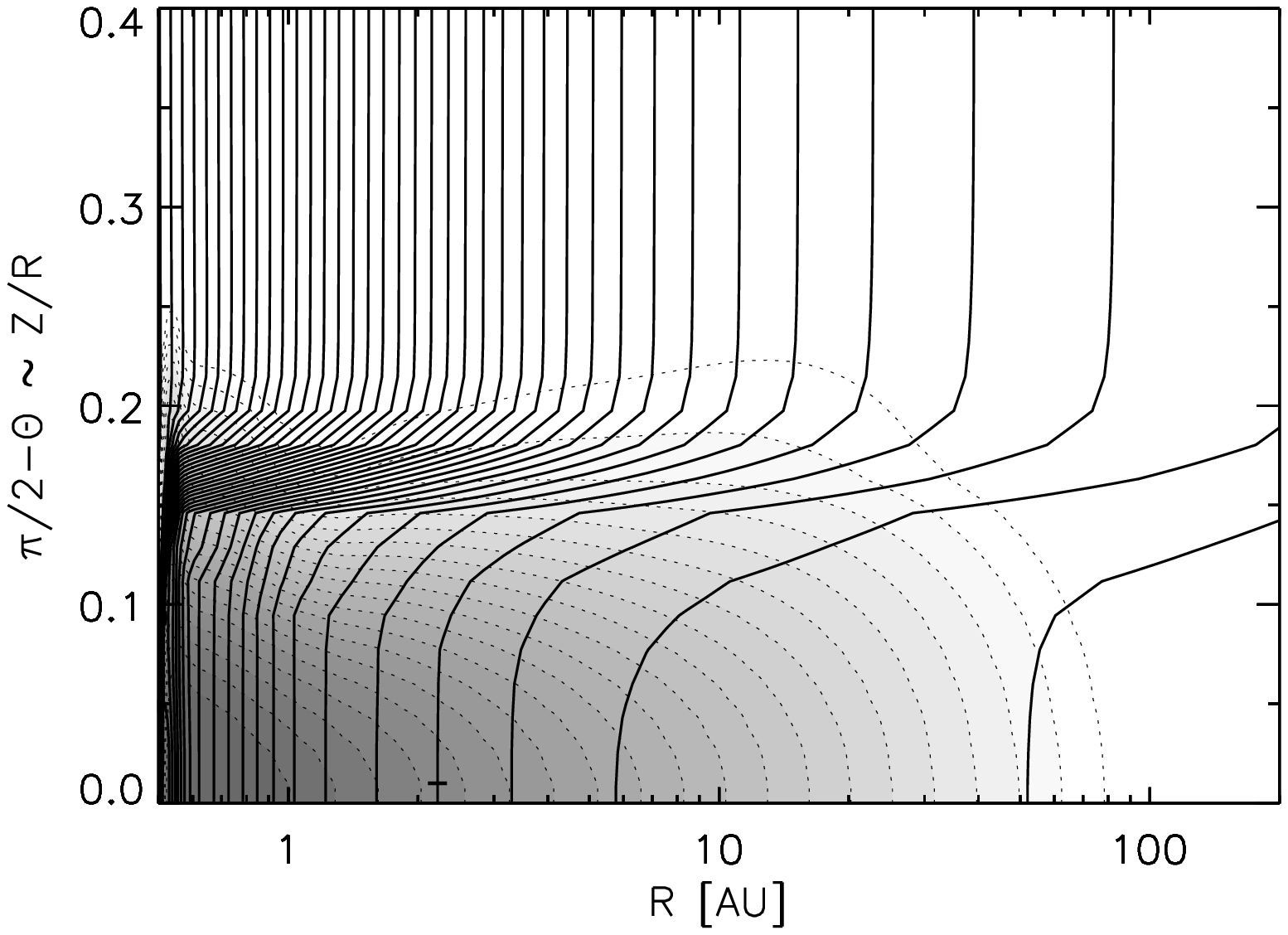}
}
\caption{The temperature and density structure as a fuction of spherical
radius $r$ and polar angle $\theta$ ($\pi/2$ is the equatorial plane,
i.e.~the bottom of the figure) for model A1 (left) and A4 (right).  The
grey-scale contours (accentuated by the dotted contour lines) are
density. These are spaced logarithmically in steps of a factor 3.3. The
solid contours are temperature. The temperature contours are spaced 50 K
apart. The small tick mark on one of these contours tags the 200 K contour.}
\label{fig-a-structs}
\end{figure*}

\begin{figure*}
\ccline{
\includegraphics[width=9cm]{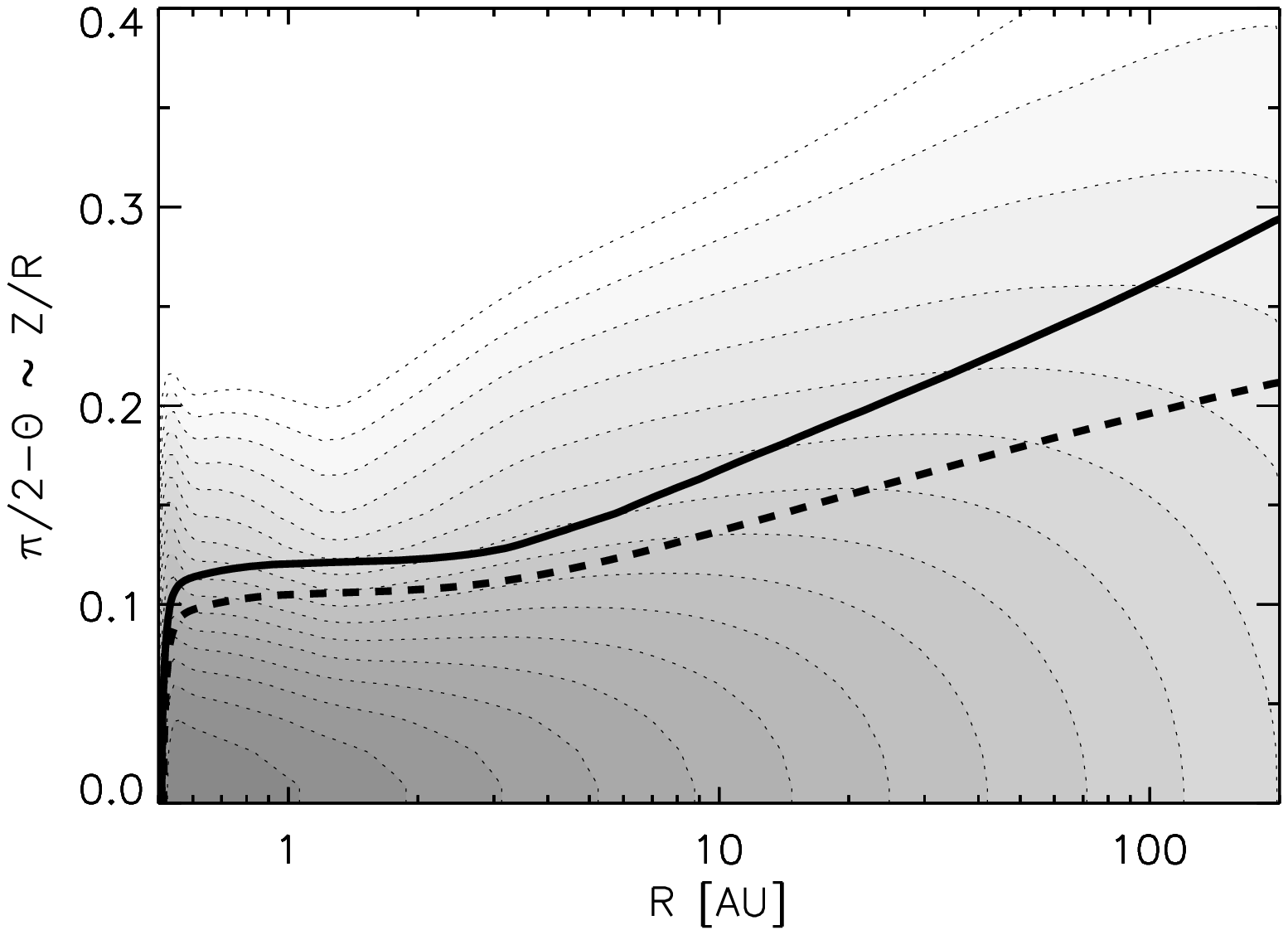}
\includegraphics[width=9cm]{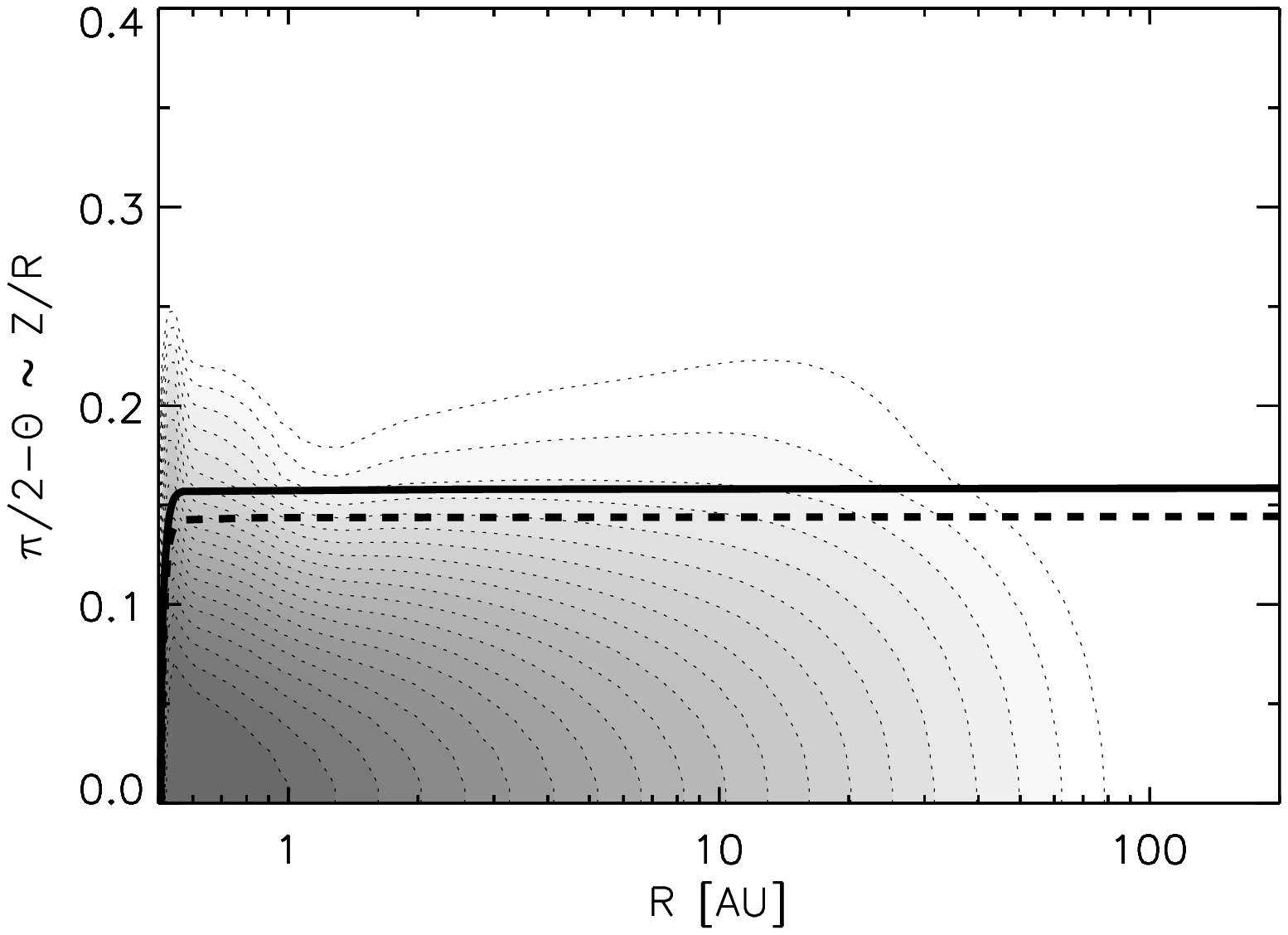}
}
\caption{As Fig.~\ref{fig-a-structs}, but instead of temperature contours,
the figure shows the surface of $\tau_{0.55\mu m,radial}=1$ (solid), and $\tau_{3\mu
m,radial}=1$ (dashed). The shape of these curves clearly show that the left
model is flared while the right model is self-shadowed. In the flared model
it is also seen that the flaring starts for real beyond about 8 AU; between
0.5 and 8 AU the curve stays at almost constant $\Theta$, typical for
the shadowed region of the disk.}
\label{fig-a-struct-tausurf}
\end{figure*}

\cdrev{To show the structure of the computed models, we use contours of
temperature and density in a plot in which the 2-dimensional spherical
coordinates $\log(R)$ and $\pi/2-\Theta$ are on the x- and y-axes
respectively. \dulrev{The midplane is at the bottom of the figure
($\theta=\pi/2$).}  These plots show the entire disk structure in a clear
way.  Lines at constant $\pi/2-\Theta$ \dulrev{(radial rays)} are horizontal
lines in this plot.  Flaring and non-flaring disks can therefore easily be
distinguished by plotting the $\tau=1$ \dulrev{location along these lines}:
flaring disks will be indicated by rising curves while non-flaring curves
will show flat curves.}  The resulting \dulrev{temperature and density}
structures for models A1 and A4 are shown in Fig.~\ref{fig-a-structs} in the
way explained above. \dulrev{In Fig.~\ref{fig-a-struct-tausurf} the same
models are shown, but this time only the density contours. Overplotted are
the $\tau=1$ surfaces at $0.55\mu$m and $3\mu$m. In these figures one can
clearly see the puffed-up structure of the inner rim in both models, in
agreement with the DDN01 model. But at large radii only model A1 has rising
density contours, while model A4 has constant or even declining density
contours. This difference is typical of the the difference between flaring
and self-shadowed disks.  The distinction is more clear in the shape of the
$\tau=1$ surfaces: for the flaring model (A1) they move upward toward larger
radius, while for the self-shadowed model (A4) they remain at constant
$\pi/2-\theta$. It is interesting to note that for the flaring model (A1)
the upward movement of the $\tau=1$ starts around about 3 AU, while the
curve remains approximately constant below 3 AU. This is the shadow of the
inner rim that covers the innermost ($<$ 3 AU) part of the disk. This
phenomenon was predicted by DDN01, and is confirmed here.}

\dulrev{The temperature contours of Fig.~\ref{fig-a-structs} are somewhat
harder to interpret. Clearly the temperature at the disk midplane is lower
than the temperature of grains above the disk's surface. This is the
reason the disk has dust features in emission. \cdrfinal{The
`kink' in the temperature profile roughly follows the $\tau=1$
surface, since $\tau=1$ defines the surface where the direct stellar
radiation is absorbed.}
}

\dulrev{Many of the qualitative structual features of these models 
are similar to the models of D02, and we refer to that paper for a deeper
discussion and interpretation of the temperature and density contours
shown here.
}

\begin{figure*}
\ccline{
\includegraphics[width=9cm]{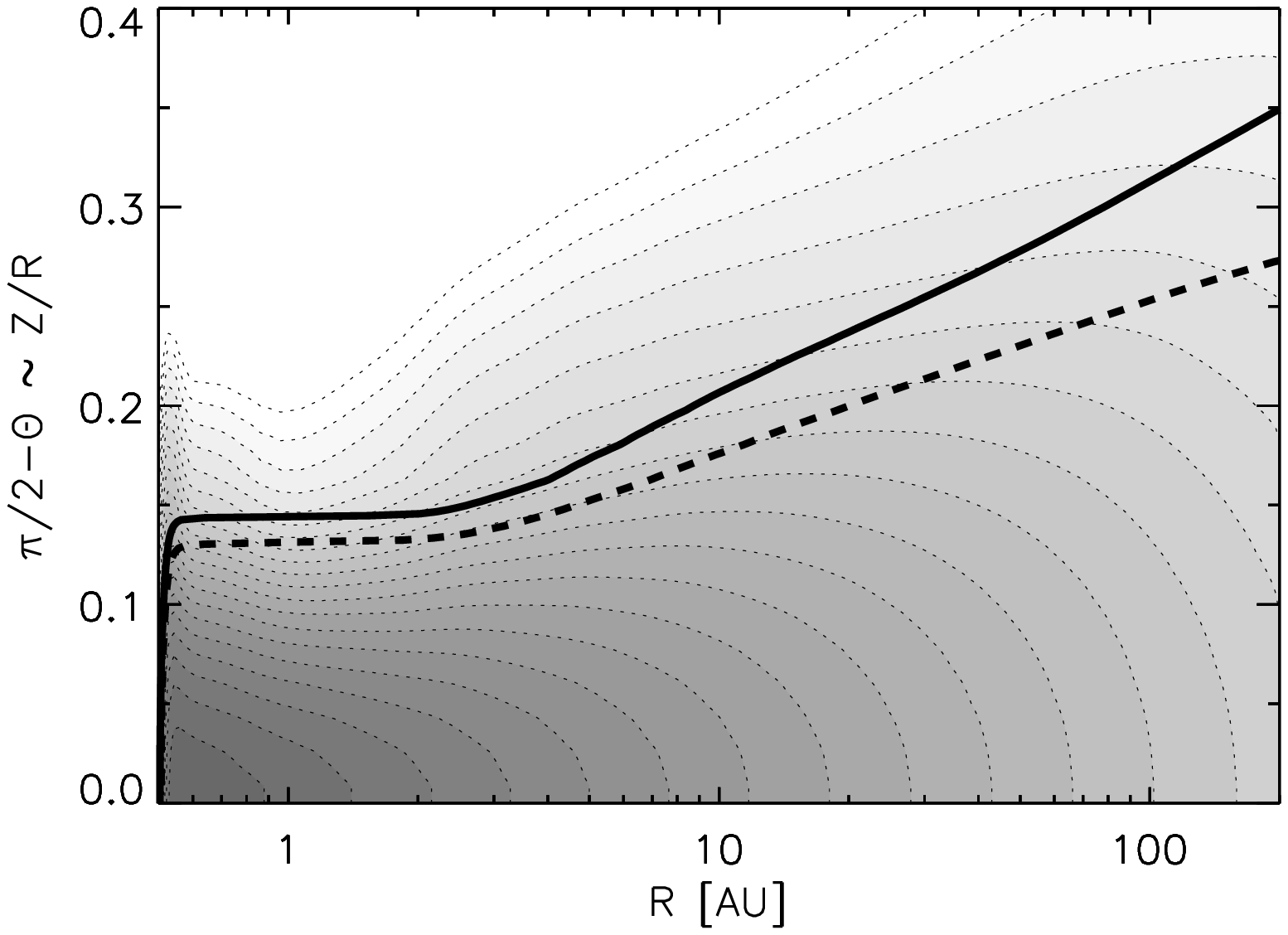}
\hspace{1em}
\includegraphics[width=9cm]{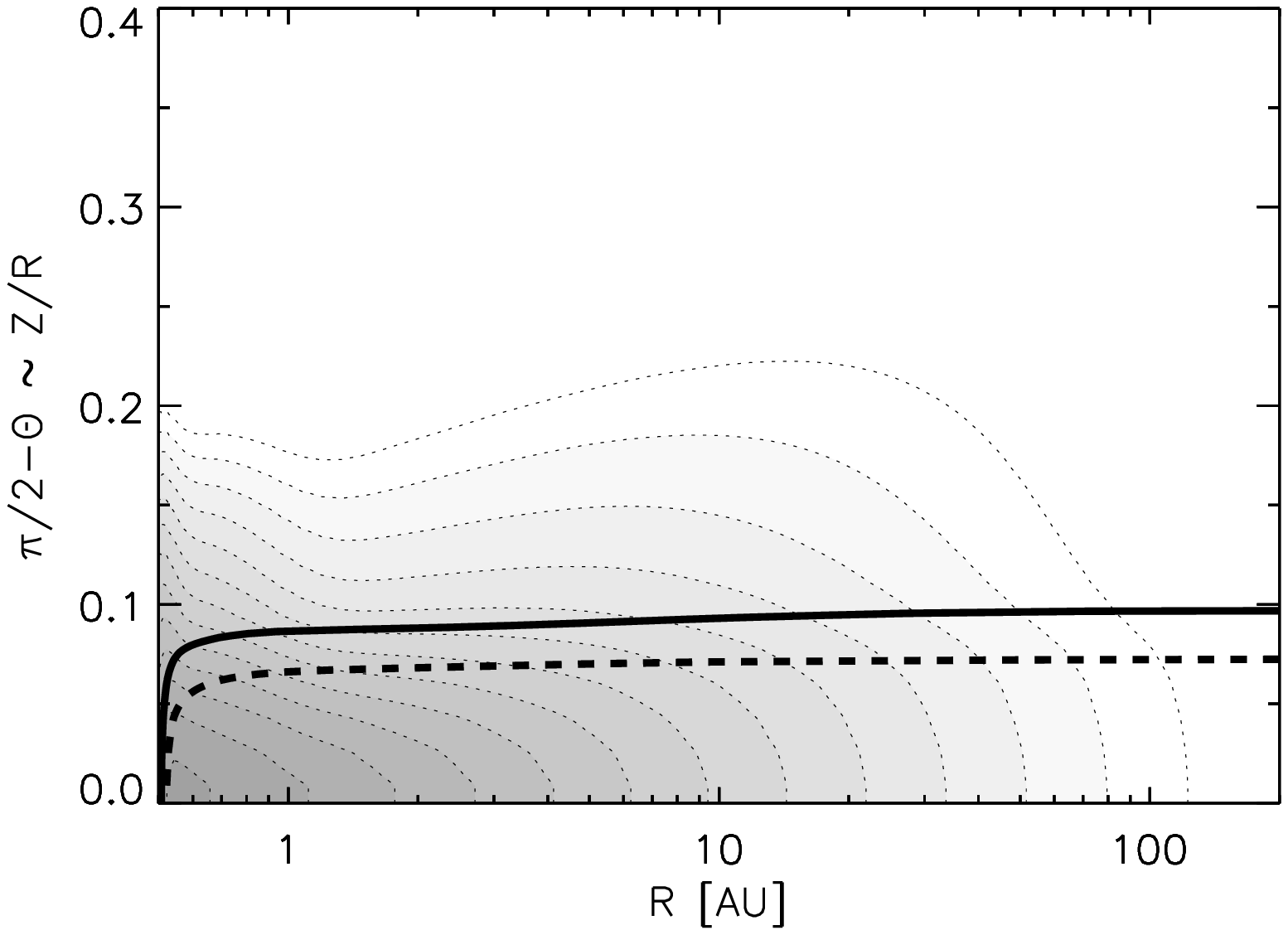}
}
\caption{As Fig.~\ref{fig-a-struct-tausurf}, but this time for 
model B1 (left) and B5 (right).}
\label{fig-b-struct-tausurf}
\end{figure*}

The models from series B shows similar results.
Fig.~\ref{fig-b-struct-tausurf} shows the stucture plots of models B1 and
B5.  Model B1 has the same kind of flaring shape as model A1.  As one goes
toward lower mass (B5) the disk tends to become more self-shadowed,
resembling model A4, but with a less high inner rim.  This shows that
decreasing the mass of the disk can have a similar effect as changing the
power law slope of the disk.  But the effect is less abrupt.  Model B4, for
example, is intermediate between a flaring and a self-shadowed disk, and
even model B5 is less self-shadowed than model A4 (compare the left panels
of Figs.~\ref{fig-a-struct-tausurf} and \ref{fig-b-struct-tausurf}).  This
is because in the B series both the height of the inner rim as well as the
height of the surface of the disk behind it get reduced. The reduction of
the disk surface height is faster, and that is the reason why from B1 to B6
the flaring disk turns into a self-shadowed one.

\cdrfinal{We do not show density and/or temperature structure figures for
  the BL series, since the global structure of the BL disk models is the
  same as for the B disk models. The only real difference is the presence of
  the passive midplane layer of large grains in the BL models.  These grains
  do not affect the temperature structure of the disk. The fact that the gas
  mass of the BL disks are all identical while for the B series they
  decrease from B1 to B6 (see table \ref{table-model-parameters}) has no
  effect on the structure, because the gas-to-dust ratio drops out of the
  equations for passive disks, as long as it remains a global constant of
  the model. Therefore, the temperature structure and, except for a global
  factor, the density structure of the B and BL disk models are the same. In
  the modeling procedure for the BL series we in fact use the density
  structure of the corresponding B model, add the midplane layer, do one
  more radiative transfer run with the RADMC code to compute the midplane
  grain temperature and finally create the SEDs for the BL series.}

\subsection{The SEDs of the disks}\label{sec-result-sed}
\begin{figure}
\ccline{\includegraphics[width=9cm]{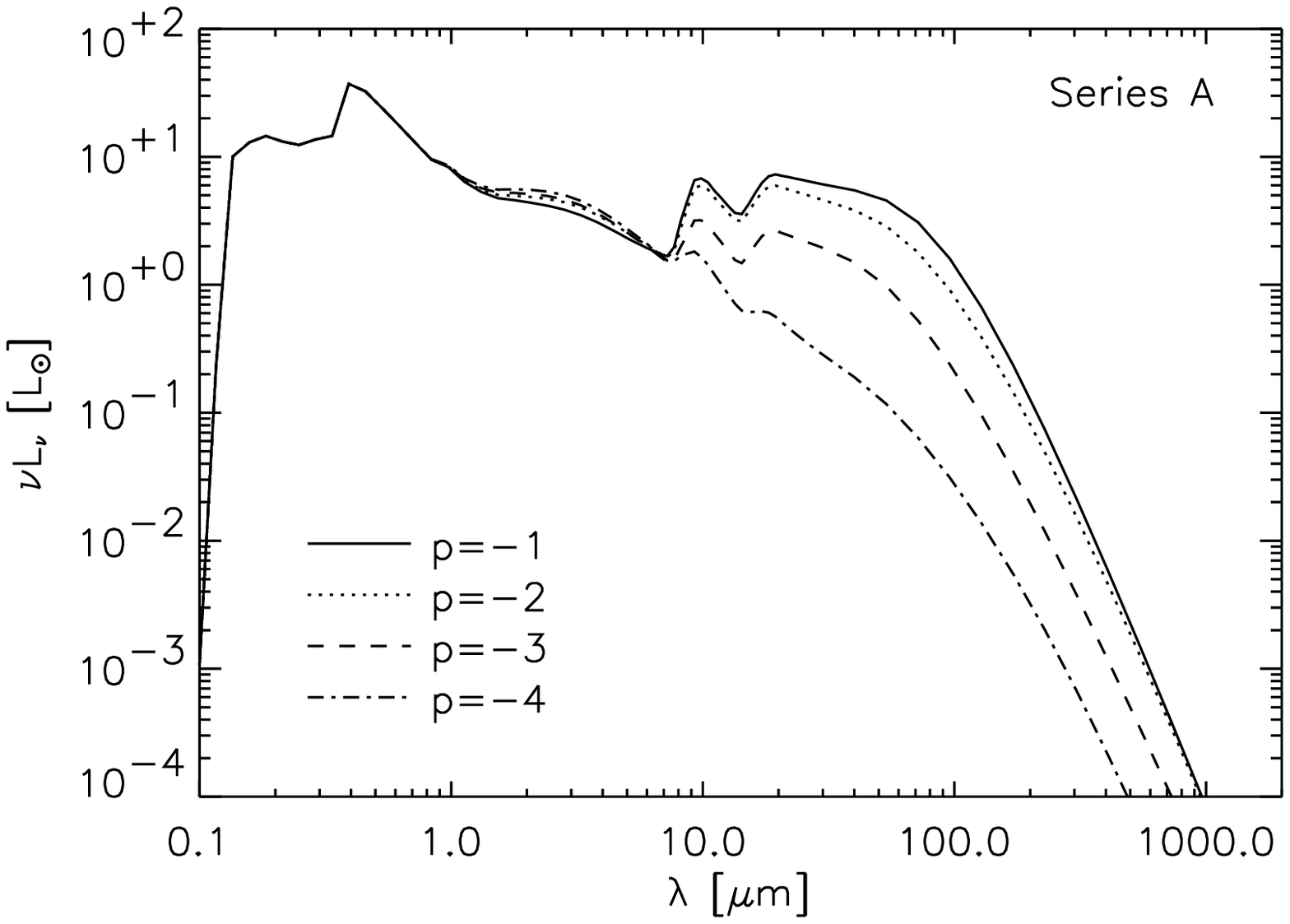}}
\caption{The SEDs of models A1...A4 plotted over each other. The SEDs are
at inclination $i=45^{o}$.}
\label{fig-A-seds}
\end{figure}

\cdrev{In Fig.~\ref{fig-A-seds} the spectral energy distributions of the
models of series A are shown. The SED varies from one that has a strong
far-IR flux to an SED that is dominated by near-IR flux. The models with a
strongly flaring outer disk (models A1,A2) are the ones that have a strong
far-IR flux. The radiation of the central star is thereby captured by the
disk at large radii, and reprocessed into far-IR emission.}  In particular
in model A1, most of the mass is in the outer regions of the disk.
\cdrev{The flaring part has a large covering fraction with respect to the
central star, and is therefore bright.  Moreover, the low density in the
inner rim region weakens the shadow effect, exposes a larger part of the
outer disk to stellar radiation and flaring, and reduces the near-IR flux.}

The models with a self-shadowed geometry (model A3,A4) turn out to have a
low far-IR flux (Fig.~\ref{fig-A-seds}). \cdrev{A significant fraction of
the disk mass resides in the inner regions, close to the rim. The inner rim
is optically thick to large heights and casts a strong shadow over the disk.
The mass in the outer regions of the disk is so low that the disk surface
stays below the shadow cast by the inner rim.  The entire disk is
non-flaring (it receives no direct radiation from the star) and falls into
the shadow, with indirect radiation from the upper parts of the inner rim as
the only remaining heating source.} Nevertheless, the silicate feature 
at 10 $\mu$m is still in emission, albeit weaker in the A4 model than in
the other models.

\begin{figure}
\ccline{\includegraphics[width=9cm]{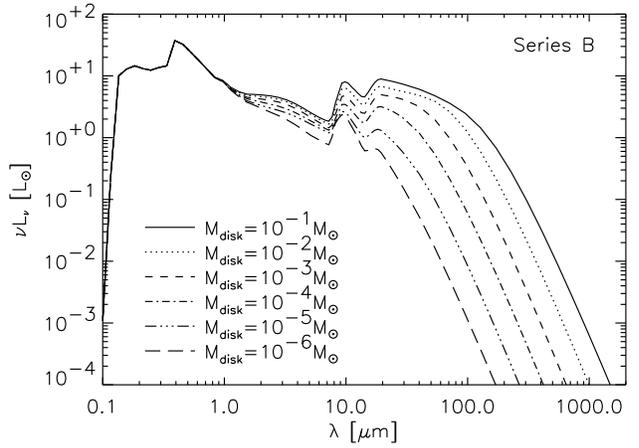}}
\caption{The SEDs of models B1...B6 plotted over each other. The SEDs are
at inclination $i=45^{o}$.}
\label{fig-B-seds}
\end{figure}

The SEDs of the B series are shown in Fig.~\ref{fig-B-seds}. The first five
models are still reasonably optically thick, and are representative for
Herbig Ae/Be stars. Models B6 starts to become optically thin even at
near-IR wavelengths, and therefore represents a transition toward optically
thin disks such as the debris disks often found around Vega-type stars. It
is clear that the flaring disk models (B1 to B3) still have a reasonably
strong far-IR flux while the self-shadowed models (B4 to B6) have a weak
far-IR flux. This is consistent with the group I/II distinction. Also, all
models have a strong 10 $\mu$ silicate feature in emission, even the
self-shadowed models (B4 to B6), which is consistent with observations. 

Interestingly, the infrared flux at different wavelengths decreases
sequentially from model B1 to B6. First the mm flux diminishes, then the
far-IR, followed by the mid-IR and finally the near-IR. The near- to far-IR
part of this trend seems to be consistent with the observed differences
between group I and group II sources. However, the (sub-)mm fluxes of the
`group II models' (B4 to B6) are much less than those of real group II
sources. In fact, the (sub-)mm fluxes of group II sources are in reality
similar to those of group I sources. Therefore, the group I/II distinction
can not be explained by merely a distinction in disk mass.

\begin{figure}
\ccline{\includegraphics[width=9cm]{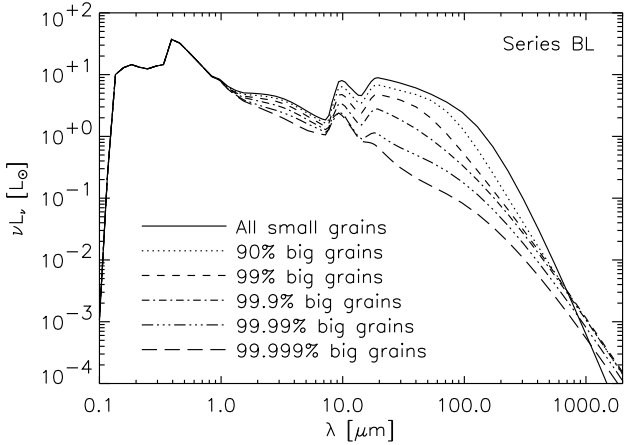}}
\caption{The SEDs of models BL1...BL6 plotted over each other. The SEDs are
at inclination $i=45^{o}$.}
\label{fig-BL-seds}
\end{figure}

The BL series is similar to the B series but the removed mass in
small grains is put into large grains located at the disk's midplane. In
Fig.~\ref{fig-BL-seds} the SEDs of these models are shown. The near- to
far-IR behavior of the BL series is the same as the B series.  But in the BL
models the midplane layer of large grains keeps the (sub-)mm flux to a
certain level, even when the far-IR flux drops from B1 to B6. Therefore,
from these results one expects no big systematic differences in the (sub-)mm
flux levels between group I and group II sources, which is consistent with
observations.

Another striking feature of the BL series is that model BL1 (which only has
small grains) has a much steeper (sub-)mm slope than the BL2 to BL6 models.
If grain growth and settling governs the transition from group I to group
II, we would therefore predict that some (though not all) group I objects
have steep (sub-)mm slopes, while all group II sources have less steep
(i.e.~more Rayleigh-Jeans-like) slopes.  This is indeed consistent with
observations: group I disks sometimes have steep slopes, indicating that
they are dominated by small grains, while group II disks generally have much
more shallow slopes indicative of large grains (Bouwman et
al.~\citeyear{bouwkotanckwat:2000}; Acke et al.~in prep.).

One thing that is not very well in agreement with observations is that in
the B and BL series, as one goes from flaring to self-shadowed, also the
flux around 10 $\mu$m is somewhat suppressed. The ratio of 10 $\mu$m to 40
$\mu$m flux does not change too much. In reality there exist many group II
sources with a much larger flux ratio between 10 $\mu$m and the far-IR than
these models predict. 

Nevertheless, the BL series, as a model of the origin of the group I versus
group II distinction, seems to be reasonably well in agreement with
observations, and may point toward an evolutionary link between these two
groups of Herbig Ae/Be stars.

\section{Discussion}
\label{sec:discussion}
\subsection{Flaring versus self-shadowed}
\label{sec:flaring-versus-self}
\begin{figure*}
\mbox{}\vspace{1em}\\
\parbox[t]{8.2cm}{
\ccline{Group I (flared disk)}
\ccline{\includegraphics[width=7.5cm]{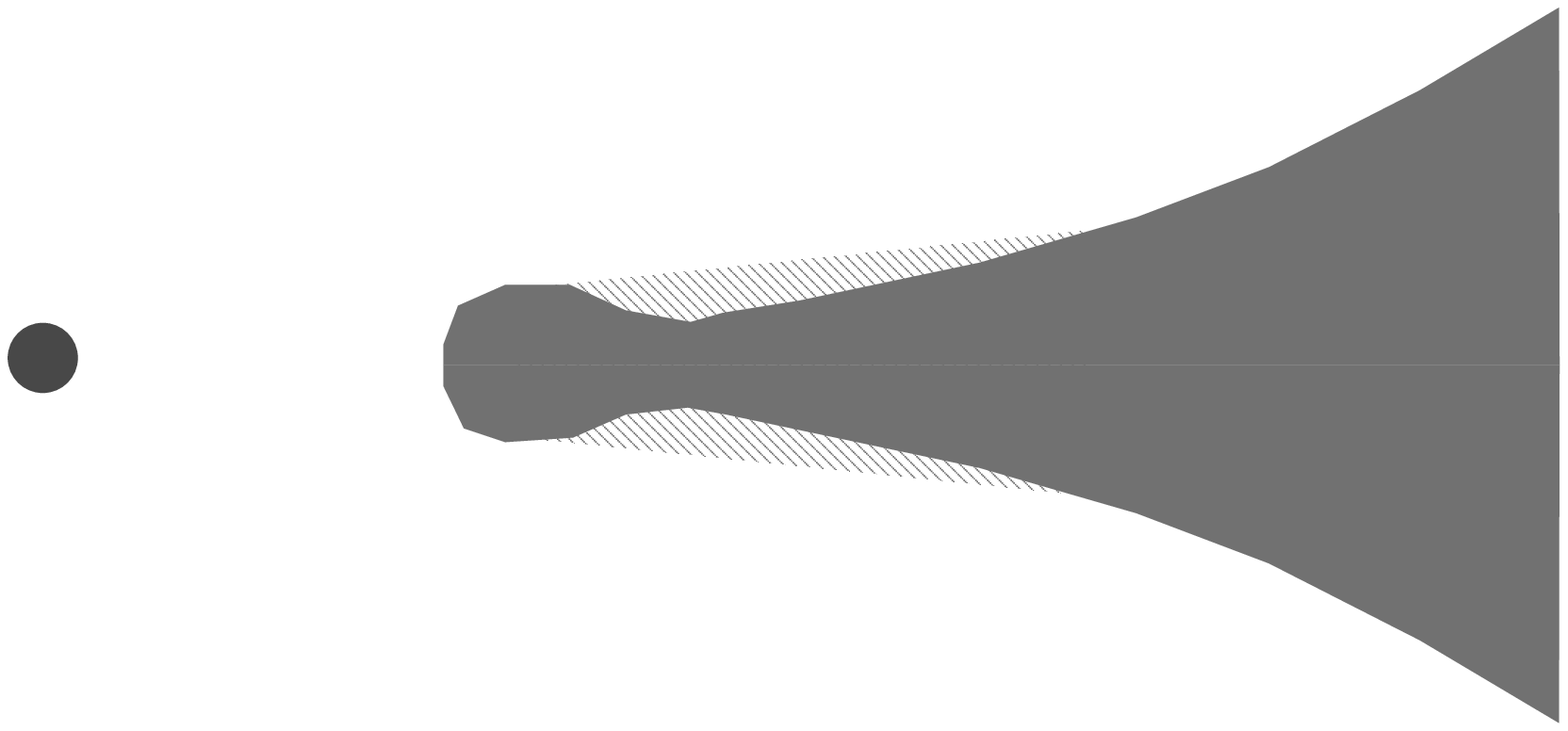}}
}
\parbox[t]{8.2cm}{
\ccline{Group II (self-shadowed disk)}
\ccline{\parbox[b]{7.5cm}{\mbox{}\vspace{1.0cm}\\
\includegraphics[width=7.5cm]{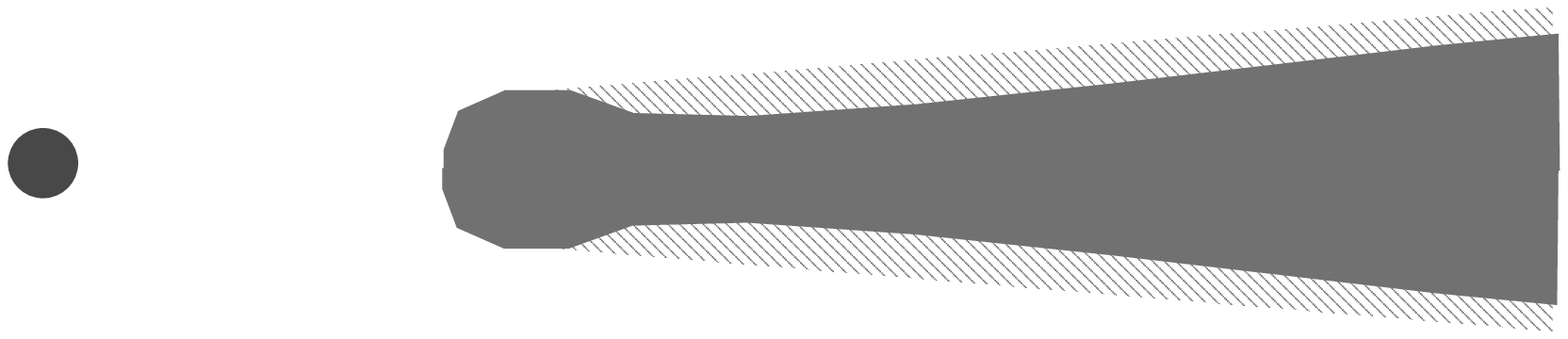}
\\\vspace{1.2cm}\mbox{}}}
}
\caption{Pictographic representation of the two disk geometries found.}
\label{fig-pictograms}
\end{figure*}

We have found two main types of solutions to the coupled 2-D continuum
radiative transfer and vertical structure problem for passive
irradiated circumstellar disks around Herbig Ae/Be stars. 
Some disks are flared from about 3 AU outwards, and other disks
are entirely self-shadowed. A pictographic representation of the two
main types of solutions is shown in Fig.~\ref{fig-pictograms}.  The
SEDs from these two types of models are very reminiscent of the two
types of SEDs observed from Herbig Ae/Be stars: those with a strong
far-IR flux and those with a weak far-IR flux.

It is not uniquely determined which parameter switches the disk from
one mode into the other. Self-shadowing is driven by two main factors:
there must be enough matter in the inner rim to produce a shadow, and
there must be too little mass at large radii to keep the disk's
surface above the shadow.  The first condition is almost always met:
already for very little matter near the inner edge the disk becomes
optically thick, and a shadow is cast. In fact, if so little matter is
present that no shadow would be cast, then the disk would presumably
be globally optically thin, and the system would not have been
classified as a Herbig Ae/Be star in the first place (it would more
resemble a Vega-type star). The second condition is not always met: as
our results show, there are situations in which the disk flares, while
in other situations it becomes self-shadowed. This depends on how high
the surface density \cdrfinal{(more accurately: opacity)} is at those large
radii. If the surface density exceeds a certain threshold, the disk
starts to flare.  This threshold, however, depends on radius and on
the hight of the shadow.

If we view these results in terms of the model parameters $M_\mathrm{disk}$
and $p$, we can see that there are two ways of arriving at the different
solutions.  For a given disk mass, the surface density powerlaw will
determine the surface densities in the outer disk.  Self-shadowed solutions
are then favored by steep powerlaws, which make the rim higher and reduce
the surface densities in the outer regions.  If on the other hand we keep
the surface density powerlaw fixed, and vary the mass, then self-shadowed
disks will be favored by low overall disk masses.

\subsection{Dust settling as the cause of self-shadowing}
\dulrev{ Another possible way of turning a flaring disk into a self-shadowed
disk is dust settling. In this paper we show that the self-shadowing
geometry can be found even without invoking dust settling, but it is
clear that dust settling will take place and may influence the models. For
compact grains with a size of about 0.1 $\mu$m the settling time scale in
the outer regions of a 100 AU disk are of the order of 1 Myr. This means
that in the typical life time of a disk the dust settles closer to the
midplane.  \cdrfinal{Depending on vertical mixing processes this may
  not proceed all the way to very thin pancake-like 
disk, but the geometric thickness of the disk can be significantly
reduced.  If by that that time the shadow of the inner rim is still
present, such disks would naturally become self-shadowed.} }

\dulrev{We leave open the question what is more important to produce a
self-shadowed disk: disk parameters, or the effect of dust settling. In a
forthcoming paper we will model the process of dust settling in detail
(Dullemond \& Dominik in prep.).}

\subsection{Shadowing and the silicate feature}\label{subsec-shad-sil}
\begin{figure}
\ccline{\includegraphics[width=7cm]{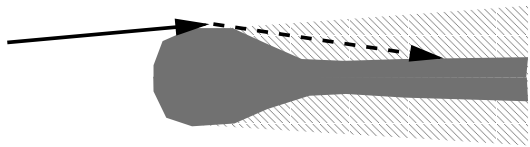}}
\caption{Pictographic representation of how the shadowed regions of
the disk can be irradiated in an indirect way through thermal emission
of (and scattering off) the upper part of the puffed-up inner rim.
Note that the geometry is exaggerated in this pictogram.}
\label{fig-indirect-irrad}
\end{figure}

The fact that in T Tauri stars and Herbig Ae/Be stars the dust features are
generally seen in {\em emission} has been explained by earlier models
(Calvet \citeyear{calvetpatino:1991}, CG97 etc) as arising from the
superheated surface layer created by the {\em direct} irradiation of the
disks's surface by the central star. Naively one would therefore expect that
the self-shadowed disks found in this paper would not have emission
features, as their surfaces do not receive direct stellar radiation. Yet, as
can be seen in Figs.~\ref{fig-A-seds}, \ref{fig-B-seds} and
\ref{fig-BL-seds}, even the self-shadowed disks have their silicate feature
strongly in emission. There are several reasons for this. First of all,
stellar photons can reach the shadowed region by indirect means
(Fig.~\ref{fig-indirect-irrad}). For instance, photons scattering off the
upper parts of the puffed-up inner rim may get diverted into the shadowed
region, and the thermal near-infrared emission from this upper part of the
rim may also irradiate the shadowed region. In fact in the models presented
here only the latter is at work, since we have neglected scattering.

Another way by which the shadowed region can be heated occurs when the
shadowing is imperfect. The A4 model has an almost perfect shadow (see
Fig.~\ref{fig-a-struct-tausurf}), while the shadowing in the B5 model is
clearly less complete (see Fig.~\ref{fig-b-struct-tausurf}). This
means that even though one cannot really speak of a flaring disk, there is
still a tail of the vertical density distribution of the disk that still
above the shadow and may therfore produce a dust emission feature.

Finally, when the disk has lost so much of it's small grain content that it
becomes optically thin at 7 $\mu$m and 13 $\mu$m, the silicate feature
naturally appears in emission. This in fact plays a role in models B5 and
B6.

The indirect heating of the shadowed regions of the disk are crucial, not
only for the silicate emission feature, but also for the far-IR flux. If the
shadowing were so perfect that no indirect heating were possible, the
self-shadowed disks outside of the inner rim would be essentially cold
(meaning that they would have the temperature of the surrounding molecular
cloud). They would therefore barely produce far-IR flux at all. This would
be inconsistent with observations of group II sources which clearly do show
some far-IR flux, albeit much weaker than that of group I sources.  The
indirect heating of these self-shadowed regions is responsible for this
weak-but-non-negligible far-IR flux, and it is reponsible for the fact that
our self-shadowed disks of the BL-series have SEDs that are so similar to
existing group II sources. 

\subsection{Extended emission: PAH features and scattering images}

\cdrfinal{An important way to study the structure of circumstellar disks in
general and self-shadowing in particular is through resolved images.  In the
sub-mm wavelength regions, all models discussed in this paper will be
extended - even though later models of series A will look more compact
because most of the disk mass has been moved to the inner regions.  A much
better test for self-shadowing will be images obtained in near and mid-IR
PAH bands, and images obtained in scattered light.  PAH emission results
from the excitation of PAHs through individual UV photons.  Scattered light
emerges from the disk when stellar photons in the optical and near-IR are
scattered off the disk surface into the line of sight. Both processes are
not modeled in the current paper, so we cannot quantitatively predict
observations.  However, since both directly track the illumination of the
disk surface with stellar photons (or, on a much lower level, stellar
photons scattered towards the disk in the upper regions of the rim), we can
make qualitative predictions.  If a disk is self-shadowed, both PAH emission
and scattered light should be reduced dramatically in the outer disk region,
much more so than in a disk which is still flaring but merely has a reduced
surface height.  In the latter case, the expected reduction factor in the
intensity of both PAH and scattered light is only a few while in fully
self-shadowed disks it should be orders of magnitude.  We note that the
models BL4\ldots BL6 are not perfectly self-shadowed - \dulfinal{a 
tenuous tail of the vertical density distribution of the disk still reaches
out of the shadow}.  However, even a small amount of dust settling would be
enough to move the available scatterers fully into the shadow.  Possible
tests for the effect of shadowing in disks therefore include correlating the
SED type (group I versus group II) with PAH emission and scattered light
images.  In group II sources, the PAH emission should be much weaker or even
absent - if present it should be compact.  Scattered light images of group
II sources should be very dim or even completely dark, \dulfinal{except at
the inner rim}.  Full studies of these effects are underway, with first
results indicating that indeed, group II sources are usually compact (van
Boekel et al, 2003) or weak in PAH emission (Acke et al, in preparation).
Sources detected so far in scattered light are consistently group I
(e.g.~Grady et al.~\citeyear{gradypolom:2001}; Grady et
al.~\citeyear{gradydevine:2000}; Danks et al.~\citeyear{danksvieira:2001}).}

\section{Conclusions}
In this paper we systematically investigated the structure and the SEDs of
passive dusty protoplanetary disks around Herbig Ae/Be stars. The models are
available in the form of ASCII tables on the internet at the following URL:
{\tt http://www.mpa-garching.mpg.de/PUBLICATIONS/DATA/\\
radtrans/flareshadow/}

The main conclusions of this paper are as follows:
\begin{enumerate}
\item The SEDs of Herbig Ae/Be stars can be quite naturally understood in
terms of a dusty circumstellar disk that is passively reprocessing the
radiation of the central star. A self-consistent 2-D axisymmetric disk model
based on 2-D continuum radiative transfer and vertical hydrostatic
equilibrium seems to be a reasonably accurate description of such a disk.
\item Two kinds of solutions are found: disks with a flaring geometry
longwards of about $3 \AU$ and those who are fully self-shadowed by the
disk's own puffed-up inner rim. The flaring disks have SEDs that seem in
agreement with observed SEDs of group I sources (in the classification
scheme of Meeus et al.~\citeyear{meeuswatersbouw:2001}), while the
self-shadowed disks have SEDs similar to those observed from group II
sources. These flared and self-shadowed disks are natural solutions of the
combined equations of radiative transfer and hydrostatics.
\item The mass distribution within the disk is one important factor in
determining whether a disk is flared or self-shadowed. The total mass of the
disk is important as well, but may conflict with the observed fact that
group II sources are not \cdrfinal{systematically} less massive than group I
sources. Keeping the mass fixed, but growing a large fraction of the grains
(by coagulation and settling) into a layer \dulrevv{of mm size grains} at
the midplane does in fact give the right trend from group I to II with a
reasonable (sub)mm flux. Moreover, this naturally leads to a less steep
(sub)mm slope for group II disks which is indeed observed. Therefore,
dust grain growth and settling can be the driving cause of an evolutionary
transition turning flaring disks (which appear as group I source) to
self-shadowed disks (appearing as group II sources).
\item All disks (both the flaring and the self-shadowed ones) have a 10
$\mu$m silicate feature in emission. This is in agreement with the known
SEDs of group I and group II sources. Only when the disk is seen edge-on,
the silicate feature can turn into absorption.
\item \dulrevv{The models predict a systematic difference in the strength
and \cdrfinal{extendedness} of PAH emission features between group I and
group II sources, with the group II sources being weaker than the group I
sources. A similar effect is predicted for resolved images of the disks in
scattered light: the group II sources being much weaker than the group I
sources, perhaps even undetectable.}
\item \dulrev{A 2D treatment of radiative transfer is essential in order to
find and treat self-shadowed disk solutions.}  
\end{enumerate}

\begin{acknowledgements}
\cdrfinal{We thank Rens Waters, Antonella Natta, Roy van Boekel, Carol Grady,
Mario van den Ancker and Bram Acke for useful discussions and
remarks.  We would also like to thank the anonymous referee for
prompting us to shorten and clarify many aspects of the paper.}
\end{acknowledgements}


\end{document}